\documentclass{PoS}
\usepackage{booktabs}
\usepackage{amsmath}

\newcommand{\Op}{\mathcal{O}} 

\title{Impact of dynamical charm quarks}

\ShortTitle{Impact of dynamical charm quarks}

\author{
   \includegraphics[width=0.2\linewidth]{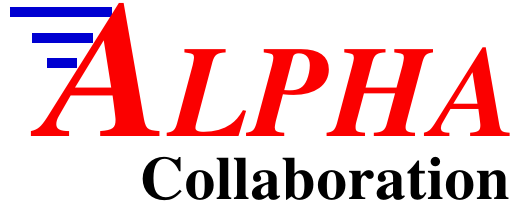}
}

\author{ \speaker{Tomasz Korzec}, Francesco~Knechtli\\
University of Wuppertal, Gau\ss str. 20, 42119 Wuppertal, Germany\\
E-mails: \email{korzec@uni-wuppertal.de},
         \email{knechtli@uni-wuppertal.de}}

\author{  Salvatore~Cal\`{i} \\
University of Wuppertal, Gau\ss str. 20, 42119 Wuppertal, Germany\\
and University of Cyprus, P.O. Box 20537, 1678 Nicosia, Cyprus\\
E-mail: \email{scali@uni-wuppertal.de}}

\author{Bj\"orn~Leder \\
Humboldt Universit\"at zu Berlin, Newtonstr. 15, 12489 Berlin, Germany \\
E-mail: \email{leder@physik.hu-berlin.de}}

\author{Graham~Moir\\
Department of Applied Mathematics and Theoretical Physics, Centre for Mathematical Sciences,
University of Cambridge, Wilberforce Road, Cambridge, CB3 0WA, UK\\
E-mail: \email{graham.moir@damtp.cam.ac.uk}}

\abstract{We compute and compare the continuum limits of several quantities in 
          QCD with and without a dynamical charm quark. We consider both low energy 
          quantities, like the hadronic scales $r_0$ and $t_0$, and high energy 
          quantities, like the charmonium masses.\\
\flushright{WUB/16-11}\\ 
\flushright{DAMTP-2016-86}\\
}

\FullConference{34th annual International Symposium on Lattice Field Theory\\
                 24-30 July 2016\\
                 University of Southampton, UK}

\begin{document}

\section{Motivation}
Many large scale QCD simulations are carried out in the 2+1 flavor theory, i.e. with light
quarks only~\cite{Lin:2008pr,Aoki:2009ix,Bietenholz:2010jr,Arthur:2012yc,Bruno:2014jqa}.
There are several good reasons for this choice. 
\begin{enumerate}
\item Lattices that are 
large enough to accommodate (nearly) physical pions, are usually too coarse to 
resolve correlation lengths associated with charm quarks. Trying to do so
leads to pronounced lattice artifacts \cite{Cho:2015ffa,Rae:2015fwa}.
\item Each additional quark flavor increases the costs of the simulation significantly. 
      Moreover, the effort of tuning the bare parameters leading to a well-defined
      chiral trajectory is greatly increased.
\item There is strong evidence that the effect of a dynamical charm quark on low energy 
      quantities requires a very high precision to be resolved~\cite{Bruno:2014ufa,Knechtli:2015lux}.
\end{enumerate}
On the other hand, charm physics becomes more and more interesting. Experiments like Belle, CLEO and BABAR
keep discovering 
new hidden and open charm-states, many of which are poorly understood. Consequently, a huge effort is made 
to explain some of these findings from first principles. But how reliable is charm physics 
on ensembles without a dynamical charm quark?

Our goal is the estimation of the effect of ``quenching'' the charm
quark, on quantities that contain valence charm quarks, e.g. on the charmonium mass spectrum.

\section{QCD with two heavy quarks}
To avoid the usual multi-scale problem, we consider a simplified version of QCD, namely 
a $SU(3)$ Yang-Mills theory coupled to two degenerate heavy quarks. This allows us to perform simulations in
relatively small volumes with very small lattice spacings. As a discretization we
use Wilson's plaquette gauge action and a clover improved doublet of twisted mass
Wilson fermions. At maximal twist, the clover term with non-perturbatively determined ~\cite{Jansen:1997nc} coefficient
$c_{\rm sw}$ is not necessary for $O(a)$ improvement of physical observables. However, it was found
that its inclusion reduces the $O(a^2)$ 
lattice artifacts, see e.g.~\cite{Dimopoulos:2009es}.

At the small lattice spacings of our simulations, critical slowing down is a major obstacle. We use
open boundary conditions in the time directions to keep auto-correlation times associated with the
topological charge manageable~\cite{Luscher:2011kk}. The boundary improvement coefficients are kept at their 
tree-level values  $c_G=1$ and $c_F=1$.

The bare coupling was chosen such that the lattice spacings cover the range $0.023~{\rm fm}\lesssim a \lesssim 0.036~{\rm fm}$.
The hopping parameter $\kappa$ was set to its critical value in order to achieve maximal twist. The critical
values were obtained from an interpolation of published data~\cite{Fritzsch:2012wq,Fritzsch:2015eka}. 
The twisted mass parameter $\mu$ was
chosen such that the RGI mass in our simulations matches that of a charm quark, more precisely, at a 
given value of the bare coupling the twisted mass parameter is
\begin{equation}
   a\mu = \frac{M}{\Lambda_{\overline{MS}}}\, Z_P(L_1)\, \frac{\bar m(L_1)}{M} \, \Lambda_{\overline{MS}} L_1 \, \frac{a}{L_1}\, ,
   \label{Eq:amu}
\end{equation}
where we set $\frac{M}{\Lambda_{\overline{MS}}}=4.87$. The pseudo-scalar renormalization constant at renormalization 
scale $L_1^{-1}$ in the Schr\"odinger 
Functional scheme, the relation between the running and the RGI mass $\bar m / M$ and the $\Lambda$
parameter of two flavor QCD in units of $L_1$ are known from~\cite{Fritzsch:2012wq}. To obtain the scale $L_1$ in 
lattice units at a particular value of the bare coupling, short interpolations were necessary.
Some of the quantities entering eq.~(\ref{Eq:amu}) have rather large errors. The errors of $Z_P$ and of the 
relative scale change
are propagated throughout our calculation, while the other amount to a change in the target value for $M/\Lambda$.

In order to quantify the impact of dynamical charm quarks, we also simulate the pure gauge theory
at values of the scales $r_0/a$, $t_0/a^2$ which are similar or larger. Table~\ref{Tab:Sim} summarizes our ensembles.

\begin{table}[t]
\centering
\begin{tabular}{c c c c c c c c}
\toprule
ID   & $\frac{T}{a}\times\left(\frac{L}{a}\right)^3$ &  $\beta$  & $\kappa$    & $a \mu$            & $r_0/a$   & $t_0/a^2$ & MDUs \\
\midrule
P    & $120\times 32^3$                              &  5.700    & 0.136698    & 0.113200           & 9.131(56) & 9.105(37) &  8592 \\
W    & $192\times 48^3$                              &  6.000    & 0.136335    & 0.072557           &14.27(15)  &22.36(13)  & 22400 \\
\midrule
qP   & $120\times 32^3$                              &  6.340    &(0.1357769)  & (0.11, 0.12, 0.13) & 9.029(80) & 9.035(30) & 20080 \\
qW   & $192\times 48^3$                              &  6.672    &(0.1353155)  & (0.07, 0.08, 0.09) &14.103(94) & 21.925(82)& 73920 \\
qX   & $192\times 64^3$                              &  6.900    &(0.1344651)  & (0.054)            &18.65(24)  & 39.43(17) &100000 \\
     &                                               &           &(0.1344597)  & (0.056)            &           &           &       \\
     &                                               &           &(0.1344540)  & (0.058)            &           &           &       \\
\bottomrule
\end{tabular}
\caption{Simulation parameters of our ensembles. The columns show the ensemble names, the lattice sizes,
the gauge coupling $\beta=6/{g_0^2}$, the hopping parameter (for maximal twist), the twisted mass parameter,
the scales $r_0/a$ and $t_0/a^2$  and the total statistics in molecular dynamics units. 
The quenched simulations need mass parameters only for the measurements.}\label{Tab:Sim}
\end{table}

\section{Strategy}
On the generated ensembles we measure the following quantities:
\subsection{Gradient flow observables}
A gradient flow equation can be solved in order to relate the simulated gauge fields to 
gauge fields at a fictitious flow-time $t$~\cite{Atiyah:1982fa}. Local operators built from gauge fields
at finite flow time do not require renormalization and have variances that remain finite in the continuum limit. 
One particularly
useful flow quantity is the flow scale $t_0$ defined by~\cite{Luscher:2010iy}
\begin{equation}
   t_0^2 \left < G^a_{\mu\nu}(t_0)G^a_{\mu\nu}(t_0) \right> = 0.3\, ,
\end{equation}
where $G_{\mu\nu}(t)$ is the field strength tensor at flow time $t$.
We use a symmetric (clover) discretization of $G_{\mu\nu}$ and the Wilson action in
the flow equation, exactly as in~\cite{Luscher:2010iy}. The action density is averaged over the time-slices far away from the 
temporal boundaries.

\subsection{Wilson loops}
Another set of useful, purely gluonic, observables are the Wilson loops. We follow~\cite{Donnellan:2010mx} and 
measure Wilson loops where the initial and final line of gauge links are smeared using up to four levels of
HYP smearing. This allows us to extract the static-quark potential $a\,V(r)$ very reliably by solving a generalized 
eigenvalue problem~\cite{Blossier:2009kd}. 
The static force $F(r) = V'(r)$ can then be used to measure the hadronic scale $r_0$ defined implicitly 
through $r^2 F(r)\bigr|_{r=r_0}= 1.65$ \cite{Sommer:1993ce} or a
renormalized coupling at scale $r^{-1}$, that can be defined by
\begin{equation}
   \alpha_{qq}(r^{-1}) = \frac{1}{C_F}r^2 F(r) \, . \label{Eq:aqq}
\end{equation}

\subsection{Meson correlation functions}
In addition to purely gluonic observables, we measure meson correlation functions\footnote{Given here in the physical basis.
In the twisted basis, the vector operators acquire an additional factor $\gamma_5$.}
\begin{eqnarray}
   C_{\Op_{\Gamma},\Op_{\Omega}}(x_0,y_0   ) &=& \langle \Op_\Gamma(x_0) \Op^\dagger_\Omega(y_0) \rangle \, \\
   \text{with} \qquad \Op_\Gamma(x_0) &=& \sum_{\vec x} \bar c(x) \Gamma c'(x)\, ,\qquad \Gamma \in \{\gamma_5, \gamma_0\gamma_5, \gamma_1, \gamma_2,\gamma_3 \} \, .
\end{eqnarray}
Here $c$ and $c'$ denote the two flavors in a twisted mass doublet.
The measurements involve stochastic time-diluted estimators with 16 $U(1)$ noise vectors.

\begin{figure}[h]
   \centering
   \includegraphics[width=0.75\linewidth]{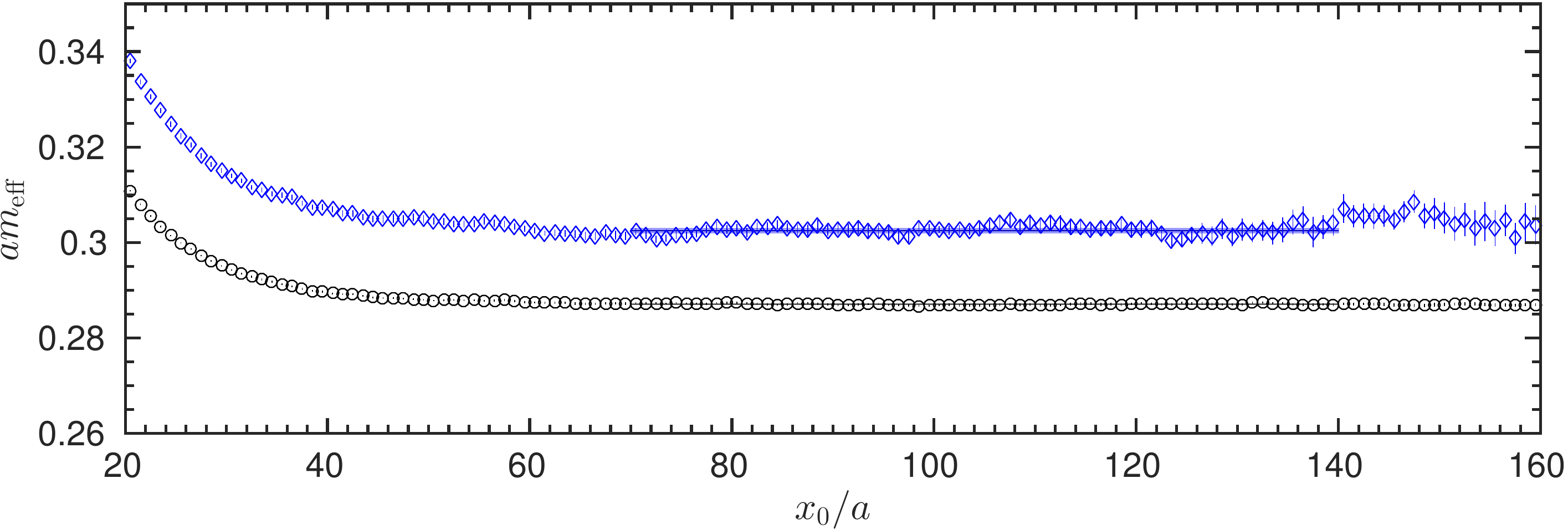}
   \caption{Effective masses and plateaux averages of the ground states in the pseudo-scalar (circles) and the 
            vector (diamonds) channel on the ensemble qX.}
   \label{Fig:effmass}
\end{figure}

The ground state energy $am_\Op$ in a channel determined by the choice of $\Op$ is then 
given by the weighted plateau average of the effective mass
\begin{equation}
   a m^{\rm eff}_{\Op}(x_0) = \log\left[ \frac{C_{\Op\Op}(a,x_0)}{C_{\Op\Op}(a,x_0+a)} \right] \, .
\end{equation}
Of particular interest will be the pseudo-scalar and vector mass
$m_P = m_{\Op_{\gamma_5}}$, $m_V = m_{\Op_{\gamma_1}} = m_{\Op_{\gamma_2}} = m_{\Op_{\gamma_3}}$.
Around the charm quark mass both pseudo-scalar and vector correlators can be measured very precisely
up to large distances between source and sink. Figure \ref{Fig:effmass} shows the effective masses and
the plateaux averages for the case of the qX ensemble.

In addition to the meson masses, we monitor the PCAC mass, to make sure that we are close enough to 
maximal twist on all ensembles.

\subsection{Quenched measurements}
\begin{figure}[h]
   \includegraphics[width=0.45\linewidth]{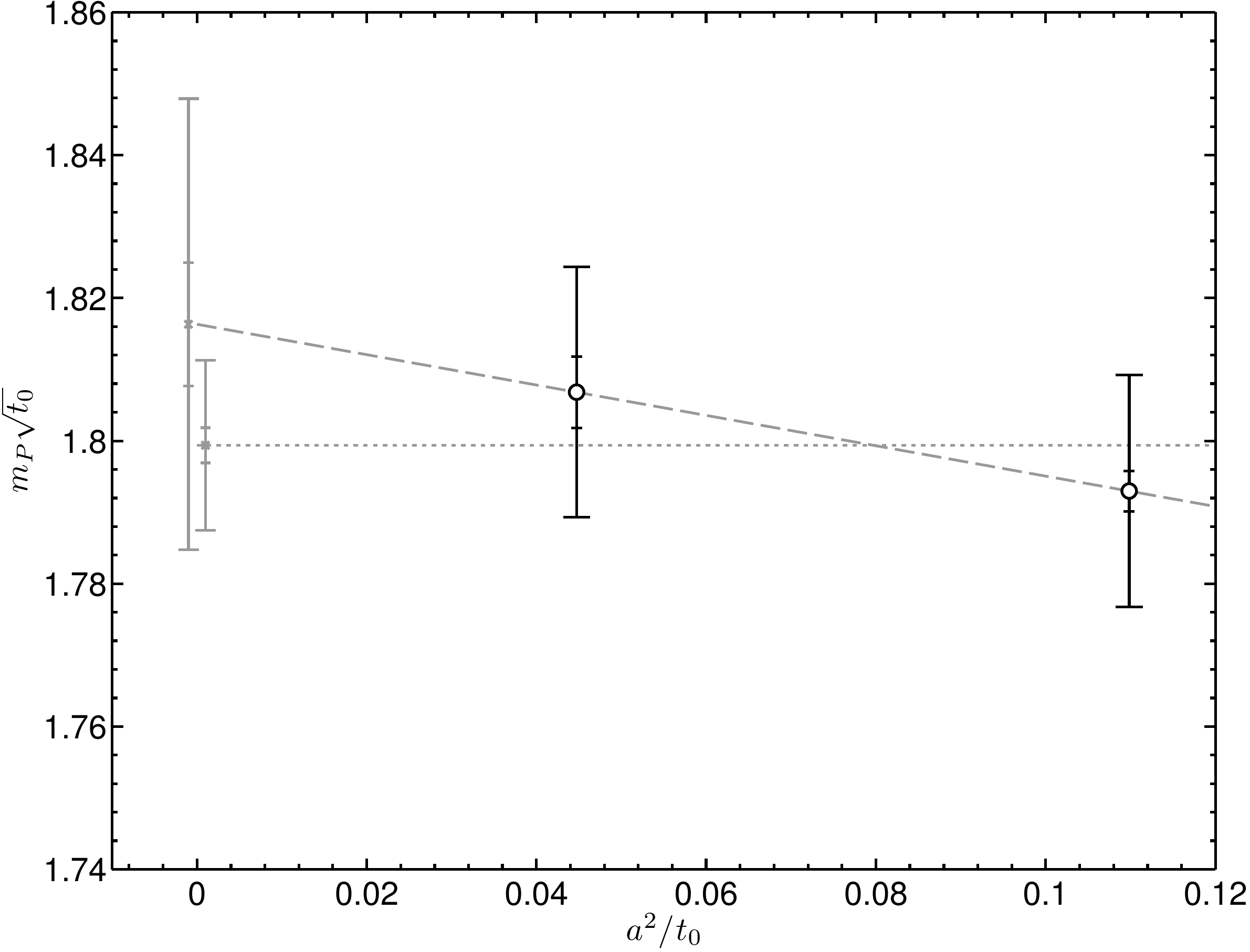}\hfill \includegraphics[width=0.45\linewidth]{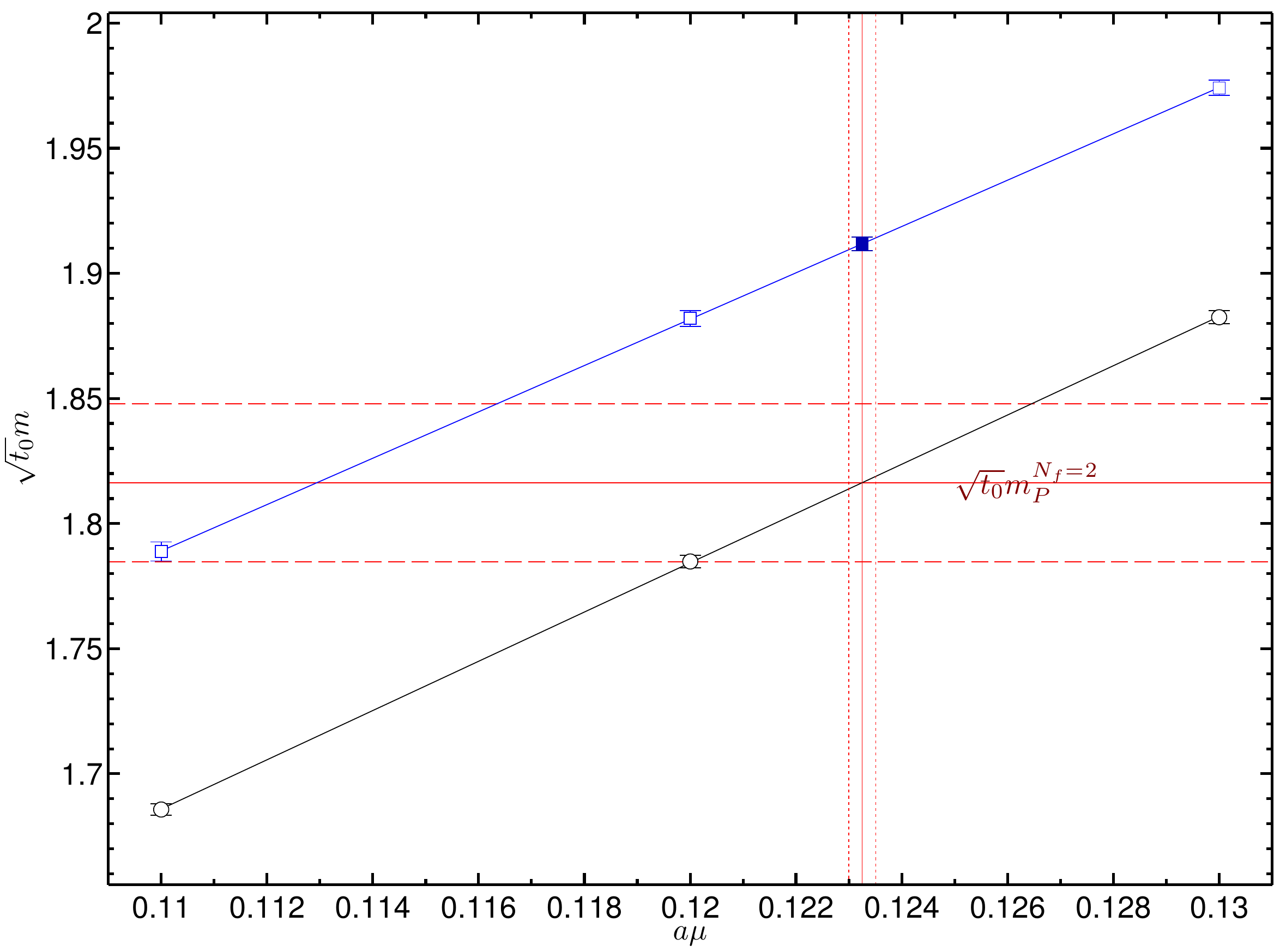}
   \caption{The left panel shows the continuum extrapolation of the pseudo-scalar mass in units of $\sqrt{t_0}$ 
            on the $N_f=2$ ensembles. Extrapolations both constant and linear in $a^2$ are shown. The continuum result from
            the linear extrapolation is used to define the tuning point for the quenched ensembles. 
            The smaller error bars are statistical errors, while the complete errors contain contributions from 
            uncertainties in $a\mu$.
            The right panel
            shows the interpolation of the measured pseudo-scalar masses (open circles) on the $N_f=0$ ensemble $qP$.
            The horizontal lines depict the tuning point and its error. The vertical lines are the resulting interpolated 
            twisted mass parameter $a\mu^*$ and its statistical error. 
            The measured vector meson masses (open squares) can then be interpolated to the
            tuning point, resulting in the solid square point. The depicted error does not contain the uncertainty in 
            $a\mu^*$, which however is taken into account in our final results.}
   \label{Fig:muinterp}
\end{figure}

In the $N_f=0$ calculations the mass and twisted mass parameters $\kappa$ and $\mu$ are required only
for the computation of observables with valence quarks. The two conditions that fix these bare parameters are
that we want to be at maximal twist $\leftrightarrow m_{PCAC}=0$, and that we want to be at
the same pseudo-scalar mass as in the dynamical simulations $\leftrightarrow \sqrt{t_0} m_P = [\sqrt{t_0} m_P]^{N_f=2}_{\rm cont.}$\, .
The critical hopping parameters were obtained from an interpolation of values in~\cite{Luscher:1996ug}, where also 
$c_{\rm sw}$ was determined non-perturbatively. The measurements
were carried out at three values of $\mu$, so that a safe interpolation to the tuning-point could be performed.
The PCAC masses turned out to be too large on the qX ensemble,
so the parameter $\kappa$ was re-tuned (for each $\mu$ value separately).
Figure~\ref{Fig:muinterp} shows the interpolation procedure on the ensemble qP. The situation is very similar 
for the two finer lattices.
On the quenched ensembles qP, qW and qX the twisted mass parameters $a\mu^*$ that lead to $\sqrt{t_0} m_P = 1.816(32)$ are
$a \mu^* = 0.1233(03)(32)$, $a \mu^* = 0.0781(03)(20)$ and $a \mu^* = 0.0581(02)(15)$ respectively. The source of the
first error is the statistical precision of $\sqrt{t_0}m_P$ on the quenched ensembles. The second error is due to
the uncertainty of the tuning goal.

\section{Results and conclusions}
\begin{figure}
 \includegraphics[width=0.45\linewidth]{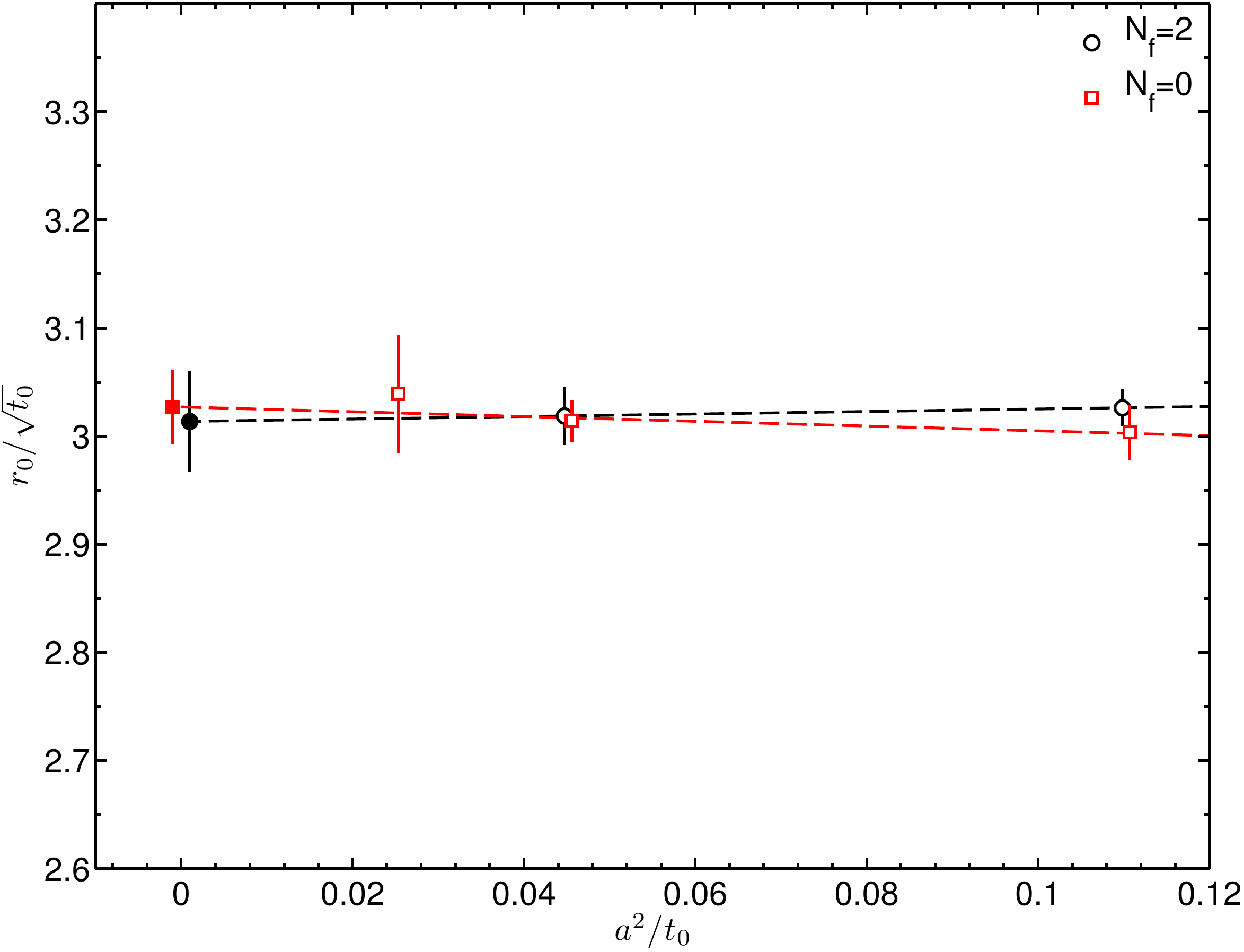}\hfill \includegraphics[width=0.45\linewidth]{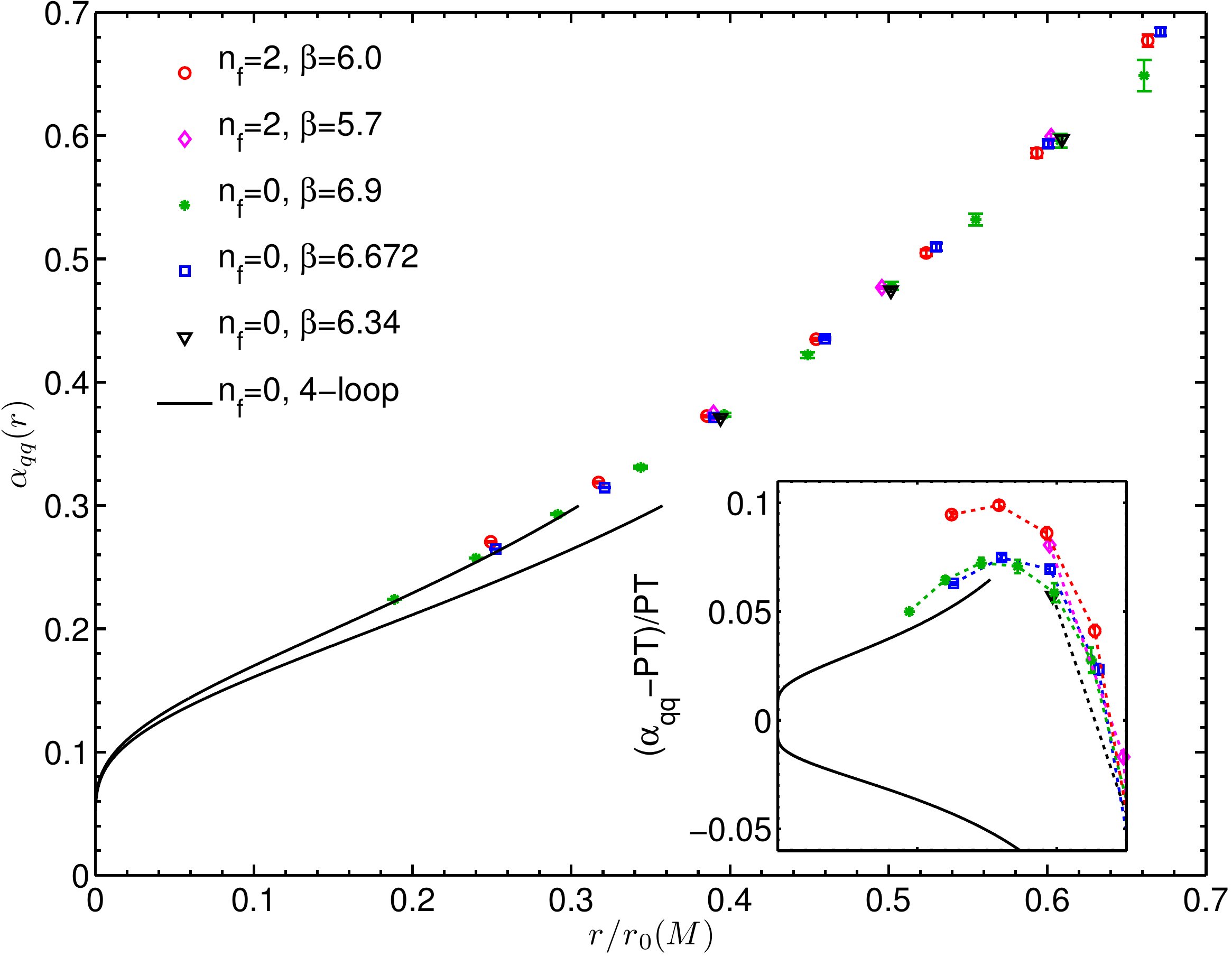}
 \caption{Comparison of purely gluonic observables in $N_f=2$ and $N_f=0$ theories. In the left panel
 the continuum extrapolations of a dimensionless ratio of scales in the two theories is shown. The continuum 
 results (solid markers) are compatible with each other. In the right panel the results for the renormalized 
 coupling $\alpha_{qq}$ are shown for all our ensembles. The solid lines stem from four loop 
 perturbation theory in the $N_f=0$ theory. Their spread is due to the uncertainty in $\Lambda$.
 In the magnified high energy part a clear dependence of $\alpha_{qq}$ on the number of flavors
 can be observed.}
 \label{Fig:r0aqq}
\end{figure}

In low energy observables we cannot resolve an effect of the dynamical heavy quark. This is shown in 
figure~\ref{Fig:r0aqq} for the ratio of hadronic scales $\frac{r_0}{\sqrt{t_0}}$, for which an effect 
of $\lesssim 0.3\%$ was predicted in ~\cite{Bruno:2014ufa}.
At our current precision of $1.5\%$ it would be a surprise to see a significant difference.

Another example is given by $\alpha_{qq}(r^{-1})$ at low energies. Here however a difference that cannot be explained by 
lattice artifacts is observed, once the renormalization scale rises to about 2~GeV.

\begin{figure}[h]
\includegraphics[width=0.45\linewidth]{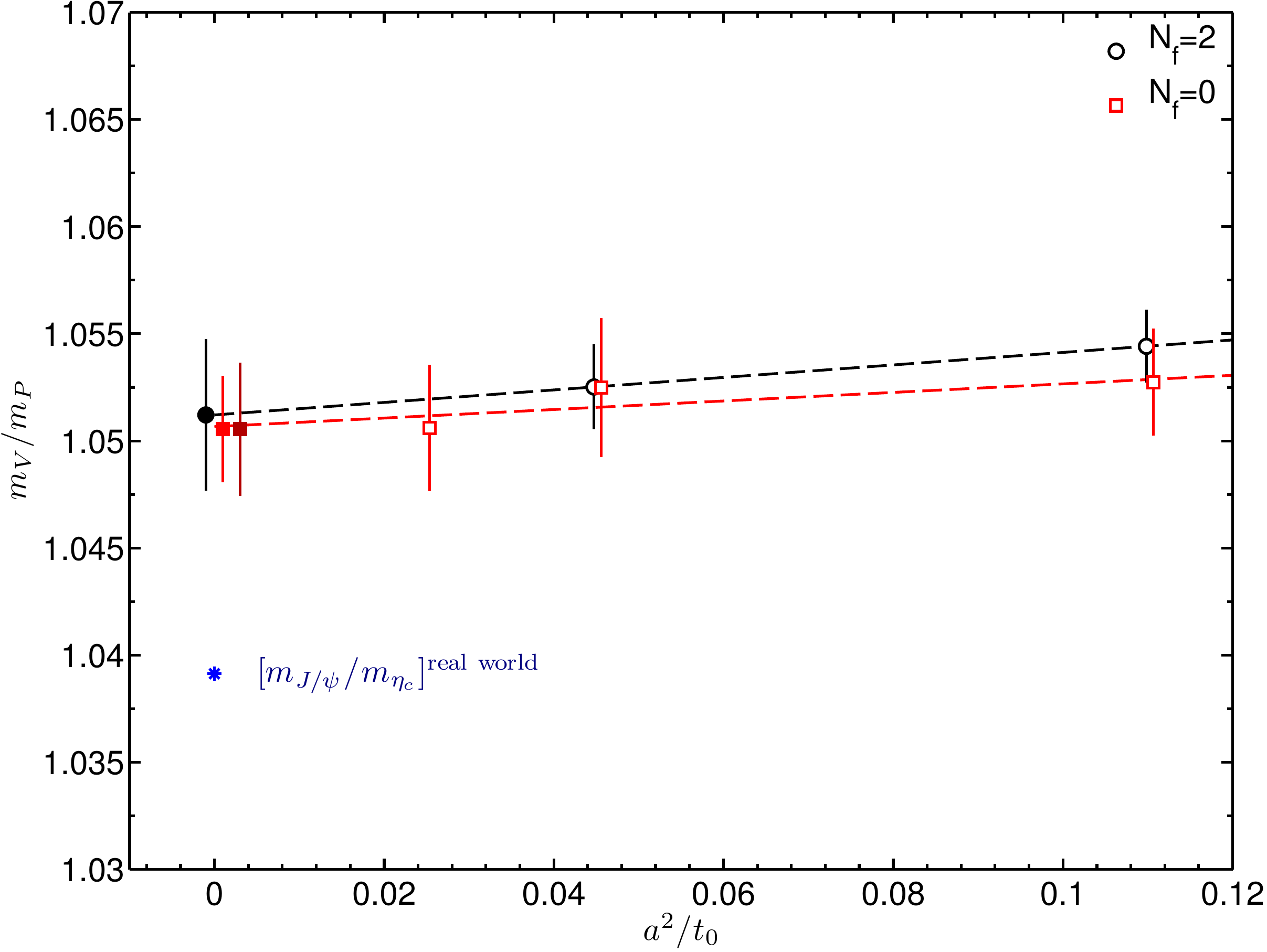}\hfill \includegraphics[width=0.45\linewidth]{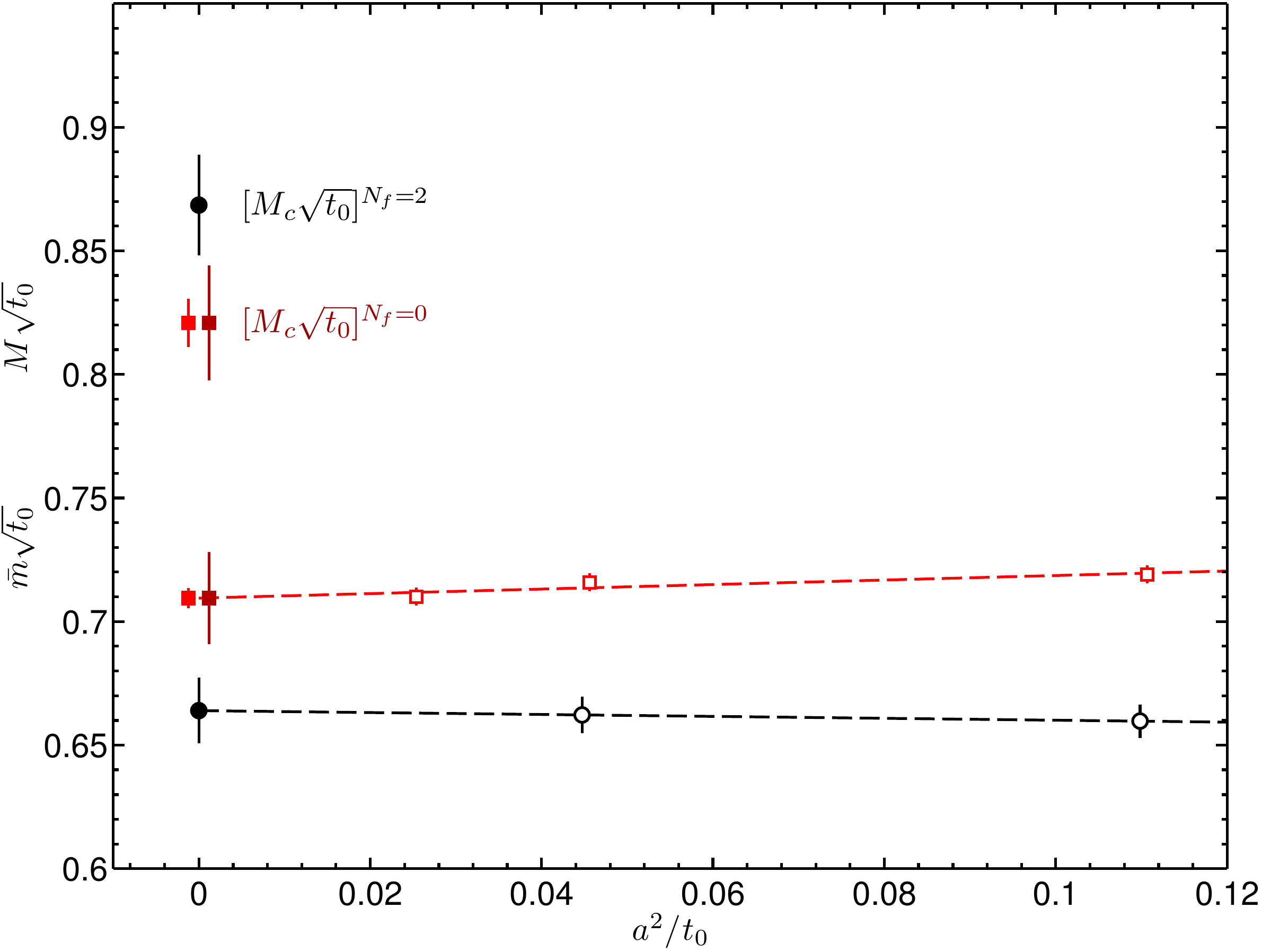}
\caption{The left panel shows the continuum extrapolation of the ratio of vector to pseudo-scalar meson mass.
The continuum extrapolated values with errors are shown. The measured ratio of the corresponding charmed
mesons is shown as well. Light sea quarks, disconnected contributions and electromagnetism are
presumably responsible for the $1\%$ deviation to our number.
The right panel shows a continuum extrapolation of the running masses. Their continuum values are
translated to RGI masses which are shown as well.
In the $N_f=0$ case, the smaller error neglects
uncertainties in $a\mu^*$, while the larger includes them.}
\label{Fig:mV}
\end{figure}

Two cases of observables involving valence charm quarks are shown in figure~\ref{Fig:mV}.
The first is the ratio of vector over pseudo-scalar meson mass. This quantity is particularly precise, 
because the two mesons have a similar dependence on the bare mass and errors due to uncertainties in $a\mu^*$
cancel to a large extent. We find that the impact of dynamical charm content in the sea on this quantity 
is tiny and below $0.3\%$. The second is the RGI quark mass $M_c$ in units of $\sqrt{t_0}$. We first determine 
the continuum values of $\bar m\sqrt{t_0}$. The agreement/disagreement of these continuum values is meaningless, 
because the running masses $\bar m$ are not renormalized at the same scale. However, since the ratios $\bar m / M$ 
are known in the two theories~\cite{Fritzsch:2012wq, Juttner:2004tb}, both can be
translated into a RGI mass, for which a comparison makes sense. We find a deviation between the RGI masses of the
two theories of
$(5.5\pm 3.5) \%$. It would be certainly useful to further reduce the error on this number.

\section*{Acknowledgments}
We thank Rainer Sommer and Jochen Heitger for many fruitful discussions and 
Martin L\"uscher and
Stefan Schaefer for making openQCD~\cite{Luscher:2012av} available.
We gratefully acknowledge the computer resources
granted by the Gauss Centre for Supercomputing (GCS)
through the John von Neumann Institute for Computing (NIC) on the GCS share
of the supercomputer JUQUEEN at JSC,
with funding by the German Federal Ministry of Education and Research
(BMBF) and the German State Ministries for Research
of Baden-W\"urttemberg (MWK), Bayern (StMWFK) and
Nordrhein-Westfalen (MIWF).
We are further grateful for
computer time allocated for our project
on the Konrad and Gottfried computers at
the North-German Supercomputing Alliance HLRN.
This project has received funding from the European Union's Horizon 2020 research and innovation programme under 
the Marie Sklodowska-Curie grant agreement No 642069.
G.M. acknowledges support from the Herchel Smith Fund at the University of Cambridge.

\bibliographystyle{JHEP-2}
\bibliography{lat}
\end{document}